\documentstyle[11pt,cite]{article}
\newcommand{\be}{\begin{eqnarray}}

\newcommand{\ee}{\end{eqnarray}}

\title{
	\begin{flushright}
	{\normalsize
        DOE/ER/40561--251--INT96--00--120\\
	March 1996 \\}
	\end{flushright}
\bf     Small x parton distributions and initial conditions
        in ultrarelativistic nuclear collisions.
       }
\author{Raju Venugopalan \\
	{\small\it Institute for Nuclear Theory,
	University of Washington,
	Seattle, WA 98195 } \\          
       }

\date{}

\begin{document}
\setcounter{page}{0}
\maketitle
\thispagestyle{empty}
\begin{center}
{\bf Abstract}\\
\end{center}

\noindent
At the colliders RHIC and LHC, nuclei at the ultrarelativistic energies of
$100$ GeV/A and $2.7$ TeV/A will be smashed together with the hope of
creating an elusive and short-lived state of matter called the quark gluon
plasma. The initial conditions which determine the dynamical evolution of the
quark gluon matter formed in the central region after the collision depend
crucially on the small $x$ component of the nuclear wavefunction before the
collision. In this comment, we discuss recent work which argues that, for large
nuclei, weak coupling techniques in QCD can be used to calculate the
distribution of these small $x$, or wee, partons. The ramifications of this
approach for the dynamics of heavy ion collisions and the various signatures of
a quark gluon phase of matter are discussed. 

\vfill \eject

\section{Introduction}

What does a nucleus look like when it is boosted to ultrarelativistic energies?
The special theory of relativity tells us that the nucleus must contract a
distance $R/\gamma$ in direction of its motion, where $R$ is its radius and
$\gamma>>1$ is the Lorentz factor. If we increase $\gamma$
indefinitely, do we expect the longitudinal size of the nucleus to shrink to
a point? What does this statement mean in terms of the underlying parton
degrees of freedom? What happpens to its transverse size--does it approach
a constant at asymptotic energies or does it keep growing~\cite{Wu}?

With the advent of the Relativistic Heavy Ion Collider (RHIC) and the Large
Hadron Collider (LHC), which will collide large nuclei at ultrarelativistic
energies, the above questions are not merely academic but are extremely
relevant to understanding the initial conditions for these collisions. The
primary objective of heavy ion collisions at these energies is to investigate
the possible formation of a soup of quark gluon matter--often simply called the
quark gluon plasma--and a phase transition of the plasma to hadronic matter~
\cite{QM95}. The formation of the plasma and indeed the dynamics of any
subsequent phase transition to hadronic matter, will depend sensitively on
these initial conditions.

In this comment, I will discuss recent
work~\cite{LV1,LV2,LV3,LV4,LV5,Raju,KLW1,KLW2} which seeks to answer the above
questions quantitatively by addressing the problem of initial conditions for
nuclear collisions within the the framework of Quantum Chromodynamics (QCD).
The center of mass energies of the colliding nuclei at RHIC and LHC are 100 GeV
and 2.7 TeV respectively. Since these energies are far greater than the typical
energies of nuclear interactions, the appropriate degrees of freedom in
describing these relativistic nuclei must be quarks and gluons, whose
interactions are described by QCD.

For ultrarelativistic nuclear collisions at central rapidities, the properties
of quarks and gluons at very low values of $x\approx k_t/\sqrt{s}$  are
relevant. (Note that $x$ is the light cone momentum fraction of the nuclear
momentum carried by the quark or gluon, $k_t$ is its transverse momentum and
$\sqrt{s}$ is the center of mass energy). Recently, there has been renewed
interest in QCD at small $x$ because of the results of the deeply inelastic
electron proton scattering experiments for $Q^2>>\Lambda_{QCD}^2$ at HERA and
the nuclear shadowing experiments at Fermilab and CERN. For an excellent
introduction to the field, see Ref.~\cite{smallx}. The results of the HERA
experiments show a very rapid rise in parton distributions for $x<<1$ which is
explained both by the conventional (operator product expansion), leading twist
Double Leading Log approximation~\cite{Gross,BalFor} and the less conventional
BFKL equation~\cite{Lipatov, DelDuca}. However, in the asymptotic limit of
$x\rightarrow 0$, neither of these approximations are correct because they
would both violate the unitarity bound on the growth of cross sections at
asymptotic energies~\cite{Froissart}.

These explanations break down completely at very small $x$ because the parton
densities become very large and many body effects become important.
Consequences of parton ``overcrowding" are that two soft partons may recombine
to form a harder parton or a parton may be screened by a cloud of surrounding
wee partons~\cite{GLR,MueQiu}. These processes inhibit the growth of parton
distributions which saturate at some critical $x$. Indeed, these processes
become important in nuclei at larger values of $x$ than in nucleons. This may
explain the strong $A$ dependent shadowing seen in the deeply inelastic
scattering (DIS) off nuclei at Fermilab and CERN~\cite{Melanson}.

It may be argued that practically all one needs in order to determine the
dynamics after a nuclear collision are the empirical nuclear structure
functions at small $x$~\cite{Eskola}. Using the QCD factorization theorem,
products of these probabilities of finding a parton in the nucleus may then be
convolved with the elementary parton--parton cross sections to determine parton
scattering rates after the nuclear collision. However, factorization  breaks
down at small $x$ (central rapidities) and coherence effects become important.
Partons from one nucleus, at relevant transverse momentum scales, do not
resolve individual partons from the other nucleus.  As in the quantum theory of
scattering, one needs to take the overlap of the {\it wavefunctions}--or more
specifically, the small $x$ Fock component of the nuclear wavefunction to
determine the subsequent nuclear evolution.

This question about the nuclear wavefunction is best formulated on the light
cone using the method of light cone quantization~\cite{Brodsky}. One can write
down the light cone QCD Hamiltonian, which is separable into a kinetic term and
a potential term. Mueller has shown recently that for heavy quarkonia, where
the scale of the coupling constant is set by the mass of the ``onium" pair,
light cone perturbation theory can be used to construct multi--parton
eigenstates at small $x$~\cite{Mueller}. Note however, that despite many
attempts, which go under the label ``Light Front QCD", thus far the light cone
approach to {\it non--perturbative} QCD has only had limited
success~\cite{Matthias}.

We wish to argue that when the density of partons is extremely large, at very
low $x$ in a nucleon or in extremely large nuclei, the density of partons sets
the scale for the running of the coupling constant. In other words, if 
\be
\rho={1\over {\pi R^2}}{dN_{part}\over dy} >> \Lambda_{QCD}^2  \, ,
\ee
then $\alpha_S(\rho)<<1$.  Here we will discuss specifically the application of
weak coupling techniques in large nuclei at small values of $x$, with $x<<
A^{-1/3}$. An intrinsic scale in the problem is set by the quantity $\mu^2\sim
A^{1/3}$ fm$^{-2}$, which is the valence quark color charge squared per unit
area. Since it is the only scale in the problem, the coupling constant will run
as a function of this scale~\cite{LV1}.

In Section 2, we write down a partition function for the parton distributions
at small $x$ in the presence of the valence quarks which play the role of
external sources in the problem. The background field for this theory is the
non--Abelian analogue of the well known Weizs\"{a}cker--Williams field in
quantum electrodynamics~\cite{Jackson}. The parton distribution functions are
formally expressed as correlation functions of a 2-dimensional Euclidean field
theory with the effective coupling $\alpha_s \mu$. The correlation functions to
each order in $\alpha_s$, involve an infinite resummation to all orders in
$\alpha_s \mu$~\cite{LV2}. It was hoped that this classical theory would
generate a screening mass ($\propto \alpha_s\mu$) which regulates the infrared
behavior of the full theory. Recent lattice results~\cite{Rajiv} suggest
however that the classical problem may not be well defined in the infrared. A
possible resolution is that the mass scale for the low momentum modes is
generated at the quantum level by the high momentum modes~\cite{Kovner}. Higher
order corrections to the background field  are also discussed briefly in
Section 2.

Nuclear collisions are addressed in Section 3. Within the above picture,
nuclear collisions can be understood as the collision of two
Weizs\"{a}cker--Williams fields. Since the fields are non--Abelian, the
classical gluon field generated after the collision is obtained by solving the
non--linear Yang--Mills equations with boundary conditions specified by the
Weizs\"{a}cker--Williams field of each nucleus~\cite{KLW1,KLW2}. In the central
region of the collision, one therefore sees the highly non--perturbative (in
$\alpha_s\mu$) evolution of the Weizs\"{a}cker--Williams glue (and sea quarks).
The time scale for the dissipation of these non--linearities is on the order of
$\sim 1/\alpha_s\mu$. On time scales much larger than this time scale, the
evolution of these fields can be described by the hydrodynamic scenario put
forward by Bjorken. The quantum picture of nuclear collisions is discussed
briefly, with particular reference to the ``Onium" picture of Mueller.

In Section 4, we will briefly discuss points of commonality as well as
difference between the Weizs\"{a}cker-- Williams model and the above mentioned
models as regards their conceptual foundations as well as their predictions for
the experiments which will be performed at RHIC and LHC. These include the
parton cascade model of Geiger and Muller~\cite{GeiMul,GeiRep}, the various
string fragmentation models~\cite{String} and the ``color capacitor" models
~\cite{Eisenberg} which describe particle production in ultrarelativistic
nuclear collisions by the QCD analog of the Schwinger mechanism in quantum
electrodynamics.

Section 5 will contain our conclusions.

\section{Computing parton distributions for a large nucleus}

In this section, the problem of calculating the distributions of partons in the
nuclear wavefunction is formulated as a many body problem. We work in the
infinite momentum frame using the technique of light cone quantization. Our
gauge of choice will be light cone gauge $A^+=0$. For an excellent discussion
of the advantages of light cone quantization, we refer the reader to
Ref.~\cite{Brodsky}. In light cone quantization and light cone gauge, the
electromagnetic form factor of the hadron $F_2$, measured in deeply inelastic
scattering experiments, is simply related to parton distributions by the
relation~\cite{dkta} 
\be F_2(x,Q^2)=\langle \int^{Q^2} d^2 k_t dk^+ x
\delta(x-{k^+\over P^+})\sum_{ \lambda=\pm} a_{\lambda}^{\dagger}
a_\lambda\rangle \, . 
\ee 
We use light cone co--ordinates ($x^{\pm}=(t \pm x)/\sqrt{2}$). In the above,
$P^+$ is the momentum of the nucleus, $k^+$ and $k_t$ are the parton
longitudinal and transverse momenta respectively, $x$ is the light cone
momentum fraction, $Q^2$ is the momentum transfer squared from the projectile
and $a^{\dagger} a$ is the number density of partons in momentum space. One
only need integrate the calculated distributions up to the scale $Q^2$ to
make comparison with experiment.

\subsection{A partition function for wee partons in a large nucleus}
\vspace*{0.3cm}

In QED, the infinite momentum frame wavefunction of the system with the
external source in Eq.~(\ref{extcur}) is a coherent state~\cite{LV1}. Failing
to do the same in QCD, we compute ground state expectation values instead. The
partition function for the ground state of the low $x$ partons in the presence
of the valence quark external source is 
\be
Z=<0|e^{iTP^-}|0>=\lim_{T\rightarrow i\infty} \sum_{N} <N|e^{iTP^-}|N>_{Q} \, .
\label{qmsum}
\ee
The sum above also includes a sum over the color labels of the sources of color
charge (denoted by $Q$) generated by the valence quarks.

The light cone Hamiltonian $P^-$ (generator of translations in light cone time 
$x^{+}$), in
the presence of an external source, can be written as 
\be
	P^- & = & { 1 \over 4} F_t^2 
+ {1 \over 2} \left( \rho_F + D_t \cdot E_t 
\right) {1 \over {P^{+}}^2} \left( \rho_F + D_t \cdot E_t 
\right) \nonumber \\ 
   &+&{1\over 2}\psi^\dagger (M-\not \! P_t ) { 1 \over P^+ } (M+\not \! P_t)
\psi \, ,
\ee
where $P^-$ can be split into kinetic and potential pieces. Here $\rho_F$ is
the charge density due to the external source plus the dynamical quarks,
$A_{t}$ and $\psi$ are dynamical components of the vector and spinor fields
respectively, $E_t = \partial_{-} A_t$ is the transverse electric field and
$F_t^2$ is the transverse field strength tensor. See Ref. \cite{LV1} for more
details regarding the above expression and the conventions used.

Valence quarks are predominantly found at large values of $x$. It is therefore
reasonable to assume that they constitute the sources of the external charge
seen by the wee partons. This current takes the form 
\be
J^{\mu}_a = \delta^{\mu +}\rho_{a}(x^+,\vec{x}_{\perp}) \delta(x^-) \, .
\label{extcur}
\ee
In the gauge $A^+=0$, the static component $J^+$ is the only large component of
the valence quark current. The transverse and minus components are proportional
to $1/P^+$ and are therefore small. The current seen by the wee partons is
proportional to $\delta(x^-)$ if the valence quarks are Lorentz contracted to a
size which is much smaller than a co--moving wee parton's wavelength. This is
satisfied if $2 Rm/P^+<<1/xP^+ \Longrightarrow x<<1/Rm\sim A^{-1/3}$ where $R$
is the nuclear radius and $m$ the nucleon mass.

Evaluating the trace in the partition function for quantized sources of color
charge is difficult. We simplify the problem by resolving the transverse space
as a grid of boxes of size $d^2 x_t >> 1/ \rho_{val}$ (or equivalently, parton
transverse momenta $q_t^2<< \rho_{val}$) which contains a large number of
valence quarks and hence a large number of color charges. This allows us to
treat the sum over color configurations classically~\cite{LV1}. We average over
the color charges by introducing in the path integral representation of the
partition function the Gaussian weight 
\be
   \exp \left\{ - \frac{1}{2\mu^2} \int d^{2}x_{t}~ \rho^{2}(x) \right\} \, ,
   \label{eq:coloraverage}
\ee
where $\rho$ is the color charge density (per unit area) and the parameter
$\mu^2$ is the average color charge density squared (per unit area) in units of
the coupling constant $g$.  It can be written as 
\be
\mu^2 = \rho_{val} <Q^2> \equiv {3A\over {\pi R^2}} {4\over 3} g^2 \sim
~1.1~ A^{1/3}~ fm^{-2} \, , 
\ee
where $<Q^2>=4 g^2/3$ is the average color charge squared of a quark.

We can now write the partition function $Z$ in the Light Cone gauge $A_{-}=0$
as 
\be
	Z & = & \int~ [dA_t dA_+] [d\psi^\dagger d\psi] [d\rho] 
 \nonumber \\  
& & \exp\left( iS +ig\int d^4x A_+(x)\delta (x^-)
\rho (x)  - {1 \over {2\mu^2}} \int d^2x_t \rho^2 (0,x_t) \right) \, .
\label{funp}
\ee
Hence, the result of our manipulations is to introduce a dimensionful parameter
$\mu^2 \approx 1.1~ A^{1/3}$ fm$^{-2}$ in the theory.  If we impose current
conservation, $D_\mu J^\mu =0$ and integrate over the external sources $\rho$,
we obtain a non--local theory~\cite{LV1} containing modified propagators and
vertices.

\subsection{The classical background field of a nucleus}
\vspace*{0.3cm}

Equivalently, one can find the classical background field in the presence of
the external sources, compute correlation functions in this background field
and finally integrate over the Gaussian random sources. 

The equations of motion are
\be
	D_\mu F^{\mu \nu} = g J^\nu\, \,   ;\, \,  J^\mu_a = \delta^{+\mu}
\rho_a(x^+,x_t)  \delta(x^-) \, \,.
\ee

The background field which satisfies the above equation of motion is $A^{\pm}=
0$, $A_i(x) = \theta (x^-)\, \alpha_i (x_t)$. Also, $F_{12}=0$ and $\nabla
\cdot \alpha= g \rho(x_t)$. Because the field strength $F_{12}=0$,  $\alpha_i$
is a pure gauge: $\tau \cdot \alpha_i = -{1 \over ig} U \nabla_i U^\dagger$.
Combining the two equations results in the highly non--linear stochastic
differential equation 
\be
	{\vec \nabla} \cdot U {\vec \nabla} U^\dagger = -ig^2 \rho(x_t) \, .
\label{ugauge}
\ee
Here $\rho$ is the surface charge density associated with the current $J$.
There is no dependence on $x^-$ because we have factored out the delta
function.  The dependence on $x^+$ goes away because of the extended current
conservation law $D_{\mu}J^{\mu}= 0$ in the background field.

To compute correlation functions associated with our classical solutions,  we
must solve the above equation and integrate the rho--dependent gauge fields
over  all color orientations of the external sheet of charge. In the matrix
notation, 
\begin{eqnarray}
\langle\alpha_i^{\alpha\beta} (x_t) \alpha_j^{\alpha^\prime \beta^\prime} 
(0)\rangle &=& {-1\over g^2}\int [d\rho]
~\bigg(U(x_t) \nabla U^\dagger(x_t)\bigg)_{\rho}^{\alpha\beta} 
\bigg(U(0)\nabla U^\dagger (0)\bigg)_{\rho}^{\alpha^\prime \beta^\prime} 
\nonumber \\
&\times& \exp\left( -{1 \over {2\mu^2}}\int d^2x_t \rho^a (x_{\perp}) \rho^a
(x_{\perp}) \right) \, .
 \label{fluct} 
\end{eqnarray}

The relation between distribution functions and propagators is straightforward
and is discussed explicitly in Ref.~\cite{LV3}. 
\be
{1 \over {\pi R^2}} {{dN} \over {dxd^2k_t}} = {1\over (2\pi)^3} {1\over x} \int
d^2 x_t~ e^{ik_t x_t}~ {\rm Tr}~[\langle\alpha_i^{\alpha\beta} (x_t)
\alpha_j^{\alpha^\prime \beta^\prime} (0)\rangle]\, ,
\label{weiz1}
\ee
where the trace is over both Lorentz and color indices. 

It was believed previously that  the distribution function has the general 
form
\be
{1 \over {\pi R^2}} {{dN} \over {dx d^2k_t}} =
{{(N_c^2-1)} \over \pi^2} {1 \over x}~
{1 \over \alpha_S}H(k_t^2/\alpha_S^2 \mu^2)\, ,
\ee
where $H(k_t^2/\alpha_S^2 \mu^2)$ is a non--trivial function obtained by
explicitly solving Eq.~(11). It was further believed that in the ``weak
coupling" limit of $k_t >> g^2 \mu$, $H(y)\rightarrow 1/y$ and one obtains the
Weizs\"{a}cker--Williams result 
\be
{1 \over {\pi R^2}} {{dN} \over {dxd^2q_t}} = {{\alpha_S \mu^2(N_c^2-1)}
\over \pi^2} {1 \over {xq_t^2}}\, ,
\label{weiz2}
\ee
scaled by $\mu^2$. It was expected that the function $H(k_t^2/\alpha_s^2\mu^2)$
would have in the strong coupling region the behavior
$\alpha_s^2\mu^2/(k_t^2+M_s^2)$ where $M_{s}\sim \alpha_s\mu$ is a screening
mass which would regulate the divergence of the distribution function at small
$k_t$. If such a screening mass did exist, it would provide a simple
understanding of saturation already at the classical level.

Unfortunately, this turns out {\it not} to be the case. Unable to find an
analytic solution to Eq.~(10) we recently solved the correlation
functions numerically on the lattice using the conjugate gradient
method~\cite{Rajiv}. Our preliminary results suggest the following. Weak
(strong) coupling on the lattice holds when $0.2 g^2\mu L << (>>) 1$, where $L$
is the lattice size. In the weak coupling limit, our results indicate a
discrete transverse momentum dependence which is of the $1/k_t^2$~
Weizs\"acker--Williams form. As one increases $g^2\mu L$, there is an
additional ``transverse" contribution to the correlation function which
displays an exponential fall off determined by a ``screening mass". However,
the amplitude of this term appears to diverge as $\sim (g^2\mu L)^2$. At large
values of $g^2\mu L$, it appears that no solutions of the stochastic equations
exist and the classical theory is ill defined in the infrared.

Albeit it appears that no infrared stable screening mass is generated at the
classical level, such a screening mass may still be generated by quantum
fluctuations around the high transverse momentum modes~\cite{Kovner}. Another
possibility is that the Gaussian weight has a momentum dependence which
modifies the infrared behavior of the correlation functions. Finally, the
assumption that the source is a delta function on the light cone may be too
severe. A way to deal with this problem is suggested in very interesting recent
work by Balitsky~\cite{Balitsky}. These ideas are outside the scope of the
present work. We shall assume in the following that the computation of small
fluctuations in the background field is not modified seriously by our lattice
result (i.e., the theory can, for instance, be regularized by integrating over
the random sources with an appropriate weight).

\subsection{Quantum corrections to background field}
\vspace*{0.3cm}

We now  outline a procedure (the familiar Dyson--Schwinger
expansion~\cite{Baym}) to systematically compute quantum corrections to our
background field to all orders. The fully connected two point function is given
by the relation 
\be
\langle \langle AA\rangle\rangle_{\rho}=\langle \langle A_{cl}\rangle
\langle A_{cl}\rangle + \langle A_{q}A_{q}\rangle \rangle_{\rho} \, .
\label{eq:schdy1}
\ee
In the above, $\langle A_{cl}\rangle$ is the expectation value of the classical
field to all orders in $\hbar$. It can be expanded as $\langle A_{cl}\rangle =
A_{cl}^{(0)}+A_{cl}^{(1)}+......$ where $A_{cl}^{(0)}$ is the solution
discussed in Section 2.2. The term $\langle A_{q}A_{q}\rangle$ is the
small fluctuation Green's function computed to each order in the classical
field. The symbol $<\cdots>_{\rho}$ indicates that we have to average over the
external sources of color charge with the Gaussian weight described previously.
We will briefly discuss the computation of the small fluctuation Green's
function and how it may be used to compute the one loop correction to a) the
gluon distribution function and b) the classical field $A_{cl}^{(1)}$.

We begin by considering small fluctuations around our classical background,
$A_{cl}=A_{cl}^0 + \delta A$. Substituting this in the partition function in
Eq.~(\ref{funp}), we only keep terms  O($\delta A^2$) in the action. The small
fluctuations propagator may be computed directly from the action or directly
from the small fluctuation Yang--Mills equations, keeping terms linear in
$\delta A$, and solving the resulting eigenvalue equation. The final expression
which involves several subtle features of light cone quantization is quite
lengthy and the reader is referred to Ref.~\cite{LV4} for the details.

In Ref.~\cite{LV5}, we have used the small fluctuation Green's function in
light cone gauge to compute the one loop correction to the background field as
well as the one loop contribution to the classical field. 
The perturbative expression for the gluon distribution function to
second order in $\alpha_s$ is
\be
   \frac{1}{\pi R^{2}} \frac{dN}{dxd^{2}k_{t}}& = &
   \frac{\alpha_{s} \mu^{2} (N_{c}^{2} - 1)}{\pi^{2}}
   \frac{1}{xk^{2}_{t}}~\Bigg\{ 1 + \frac{2\alpha _s N_c}{\pi} \ln
 \left(\frac{k_t}{\alpha_s\mu}\right)~\ln \left(\frac{1}{x}\right)  \Bigg\}.
   \label{eq:gluondistfuncsec}
\ee

Equation~(\ref{eq:gluondistfuncsec}) contains both $\ln(1/x)$ and $\ln(k_t)$
corrections to the $1/(xk_t^2)$ distribution and they represent the first order
contributions to the perturbative expansion for the distribution function. In
the kinematical region of validity, these corrections are large. This signals
that in order to properly account for the perturbative corrections one has to
devise a mechanism to isolate and sum up these leading contributions. This
work is in progress~\cite{Kovner}.

The one loop corrections to the classical background field are computed as
follows. We start with the classical equations of motion $D_{\mu}F^{\mu\nu}_a =
g J^{\nu}_a$ and expand the full gluon field as $A^{\mu} = B^{\mu} + b^{\mu}$
where $B^{\mu}$ is the background (classical) field, that is $<A^{\mu}>\,
=B^{\mu}$, while $b^{\mu}$ is the fluctuation (quantum) field with $<b^{\mu}>\,
=0$. Keeping up to quadratic terms in $b^{\mu}$, one can write equations for
the $+,-$ and transverse components of the equations of motion. All the terms
involving bilinear products of $b^\mu$ in the minus and transverse components
of the equations can be shown to vanish and these equations are identical to
their classical counterparts. The equation for the $+$ component is 
\be
   -\partial_{-}\partial^{+}B_{a}^{-} -(D_i\partial^+B^i)_{a} =  gj^+_a +
g<J^{+}_{a}> \, ,
\label{eq:set} 
\ee
where the induced current in the above equation is related simply to the
small fluctuations propagator $G_{bc}^{ij}(x,y)$ by the equation
\be
 j^+_a(x)=f_{abc}<b_{b}^{i}(x)\partial^+b_{c}^{i}(x)>=
 i f_{abc}\lim_{y\rightarrow x} \frac{\partial}{\partial y^-}
   G_{bc}^{ii}(x,y)\, .
 \label{eq:term}
\ee

We find that the Fourier transform of $j^+_a(x)$ is
$g\tilde{j}^+_a(p)=\Pi_{ab}^{+i}(p)A_b^{i(0)}(p)$, where $\Pi_{ab}^{+i}(p)$
is given by 
\be
   \Pi_{ab}^{+i}(p)= g^2p^+p^i\left(\frac{5\Gamma(-\omega)}{16\pi^2}\right)
   \delta_{ab} \, ,
\ee
which is the standard expression for the $\;+i\;$ components of the
polarization operator in light cone gauge~\cite{leibbrandtreview}. We conclude
from the above that the modifications to the background field introduced by
the quantum fluctuations do not induce extra terms in the expression for the
distribution function~\cite{LV5}. This is consistent with the theorem of
Dokshitzer, Diakonov and Troyan~\cite{DDT}. The effect of quantum corrections
to the background field can be included by replacing the coupling constant $g$
by the renormalized coupling constant $g_R$ which runs as a function of
$\mu^2$. The structure of the background field at one loop remains unchanged. 

\section{Nuclear collisions of Weizs\"acker--Williams fields}
\vspace*{0.3cm}

In the previous section we discussed the properties of the
Weizs\"acker--Williams field of a single nucleus. Recently, A.~Kovner,
L.~McLerran and H.~Weigert~\cite{KLW1,KLW2} have made significant progress in
solving the {\it classical} problem of the evolution of these fields after the
nuclear collision.

Below, we outline very briefly their key results and refer the interested
reader to their papers for further details. Before the two nuclei collide (for
times $t < 0$), the Yang--Mills equations for the background field of two
nuclei on the light cone is simply  $A^{\pm}=0$ and 
\begin{eqnarray}
          A^i  =  \theta (x^-) \theta(-x^+) \alpha^i_1 (x_\perp) +
 \theta (x^+) \theta (-x^-) \alpha^i_2 (x_\perp)
\end{eqnarray}
The two dimensional vector potentials are pure gauges (as in the single nucleus
problem!) and for $t < 0$ solve $\nabla \cdot \alpha_{1,2} = g
\rho_{1,2}(x_\perp)$. The interesting aspect of this solution is that the
classical field configuration does not evolve in time for $t < 0$! This is a
consequence of the highly coherent character of the wee parton clouds in the
nuclei.

The above solution for $t < 0$ is a fairly straightforward deduction from the
single nucleus case. What is very interesting is that the above mentioned
authors find a solution to the field equations after the nuclear collision (
for $t > 0$). It is given by
\begin{eqnarray}
  A^{\pm} = \pm x^{\pm} \alpha(\tau, x_\perp)\,\,;\,\, A^i =\alpha_\perp^i 
(\tau,x_\perp) \, ,
\end{eqnarray}
where $\tau = \sqrt{t^2 - z^2} = \sqrt{2x^+x^-}$. The relation between
$A^{\pm}$ follows from the gauge condition $x^+ A^- + x^- A^+ =0$. This
solution only depends on longitudinal boost invariant variable $\tau$ and has
no dependence on the space-time rapidity variable $y = { 1 \over 2} \ln{x^+
\over x^-}$. This suggests that the parton distributions will be boost
invariant for all later times. This result therefore justifies Bjorken's
ansatz~\cite{bj1} for the subsequent hydrodynamic evolution of the system.

The above ansatz for the background field can be substituted in the Yang--Mills
equations to obtain highly non--linear equations for $\alpha(\tau,x_\perp)$ and
$\alpha_\perp^i(\tau,x_\perp)$. The detailed expressions are given in
Ref.~\cite{KLW2}. The initial conditions for the evolution of these equations
depend on the single nucleus solutions: 
\be
\alpha_\perp^i|_{\tau=0}=\alpha_1^i+\alpha_2^i\,\,\,\,;\,\,\,\,
 \alpha|_{\tau=0}=\frac{ig}
{2}~[\alpha_1^i,\alpha_2^i] \, ,
\ee
where $\alpha_{1,2}^i$ are the background fields for the two nuclei.

The Yang--Mills equations with the above boundary conditions are solved
perturbatively, order by order, by expanding the fields in powers of the
valence quark charge density $\rho$. For asymptotically large $\tau$, Kovner
et al. find that a gauge transform of the fields $\alpha$ and $\alpha_\perp^i$ 
(denoted here by $\epsilon$ and $\epsilon_\perp^i$ respectively) have the form
\begin{eqnarray}
        \epsilon^a(\tau,x_\perp) & = &
      \int {{d^2k_\perp} \over {(2\pi)^2}}
       {1 \over \sqrt{2\omega}}
      \left\{ a_1^a(\vec{k}_\perp) {1 \over \tau^{3/2}}
      e^{ik_\perp\cdot x_\perp -i\omega \tau} + C. C.
        \right\} \nonumber \\
        \vec{\epsilon^{a,i}} (\tau,x_\perp) & = &
        \int {{d^2k_\perp} \over {(2\pi)^2}}
        \kappa^i
{1 \over \sqrt{2\omega}} \left\{ a_2^a(k_\perp)
{1 \over \tau^{1/2}}
      e^{ik_\perp x_\perp-i\omega \tau} + C. C. \right\} \, ,
\end{eqnarray}
where $a_1$ and $a_2$ can be expressed in terms of the $\rho$ fields. In this
equation, the frequency is $\omega = \mid k_\perp \mid$ and the vector
$\kappa^i = \epsilon^{ij} k^j/\omega$. The notation C. C. denotes complex
conjugate.

With the above form for the fields, the expressions for the parton number
densities is straightforward. For late times, near $z=0$, one
obtains~\cite{blaimuell} 
\begin{eqnarray} {{dN} \over
{dyd^2k_\perp}} = {1\over (2\pi)^3}\sum_{i,a} \mid a_i^a(k_\perp) \mid^2
\end{eqnarray}
Averaging over the $\rho$ fields with the Gaussian weight in
Eq.~(6), one obtains the following result for the gluon
distribution at late times after the nuclear collision: 
\begin{eqnarray} {1\over {\pi R^2}}{{dN} \over
{dyd^2k_\perp}} = {16\alpha_s^3\over \pi^2}
N_c~(N_c^2-1)~{\mu^4\over k_t^4}~\ln({k_t\over {\alpha_s\mu}}) \, .
\end{eqnarray}
As suggested by the logarithm in the above equation, the transverse momentum
integrals are infrared divergent. They are cut off by a mass scale
$\alpha_s\mu$ which was believed to result from the non--perturbative behavior
of the classical fields. However, as we have discussed earlier, our recent
lattice calculations suggest instead that the theory is not well defined in
the infrared. Any such dynamically generated mass will therefore arise only
at the quantum level.

This brings up a related issue. Thus far we have only discussed the dynamical
evolution of the classical fields. What about quantum effects? One way to
include these is to do what we did for a single nucleus--look at small
fluctuations around the background field of two nuclei~\cite{sasha}. The
background field in this case is much more complicated than in the single
nucleus case and the quantum problem is significantly more difficult. Another
approach is to consider what quantum effects do to the coherence of the initial
wavepacket.

In this regard, A.~H.~Mueller's~\cite{Mueller} formulation of the low $x$
problem is relevant. He considers an ``Onium" (heavy quark--anti-quark) state
of mass $M$ for which $\alpha_S(M)\ll 1$. In weak coupling, the $n$-- gluon
component of the onium wavefunction obeys an integral equation whose kernel in
the leading logarithmic and large $N_c$ limit is precisely the BFKL kernel~
\cite{Lipatov}. The derivation relies on a picture in which the onium state
produces a cascade of soft gluons strongly ordered in their longitudinal
momentum; the $i$--th emitted gluon has a longitudinal momentum much smaller
than the $i-1$--th.

In the large $N_c$ limit, the $n$ gluons can be represented as a collection of
$n$--dipoles. Hence, in high energy onium--onium scattering, the cross section
is proportional to the product of the number of dipoles in each onium state
times the dipole--dipole scattering cross section~\cite{Mueller2}. This cross
section is given by two gluon exchange (the pomeron). More complicated
exchanges involving multi--pomeron exchange have been studied recently by
Salam~\cite{Salam}. However, despite the mathematical elegance and simple
interpretation of the onium approach, it is unclear whether it can be extended
to nuclei.

\section{Parton cascades and color capacitors}
\vspace*{0.3cm}

In this section we will briefly discuss, in relation to the model discussed in
earlier sections, some other attempts to model the initial conditions for
ultrarelativistic heavy ion collisions. They may be broadly (and somewhat
imprecisely) classified as follows: a) perturbative QCD based models which
assume the factorization theorem and incoherent multiple scattering to
construct a spacetime picture of the nuclear collision, and b)
non--perturbative models where particle production is based on string
fragmentation or pair creation in strong color fields.

Among perturbative QCD based models, the parton cascade model of Geiger and 
M\"{u}ller~\cite{GeiMul,GeiRep} has been applied extensively to study various
features of heavy ion collisions. The evolution of {\it classical} phase space
distributions of the partons is specified by a transport equation of the form
\be
\left [ {\partial\over \partial t} - \vec{v}\cdot {\partial\over\partial \vec{r}
}\right] F_a (\vec{p},\vec{r},t)=C_a (\vec{p},\vec{r},t) \, ,
\label{casc1}
\ee
where $F_a$ are the {\it classical} phase space distributions for particle type
$a$ and $C_a$ is the corresponding collision integral. The matrix elements in
the collision integral are computed from the relevant tree level diagrams in
perturbative QCD.

The initial conditions in the parton cascade model are specified at some
initial time $t=t_0$ by the distribution $F_a (\vec{p},\vec{r},t=t_0)= P_a
(\vec{p},\vec{P}) R_a (\vec{r},\vec{R})$. The momentum distribution $P_a
(\vec{p},\vec{P})=f_a (x,Q_0^2) g(p_t)$ is decomposed into an uncorrelated
product of longitudinal and transverse momentum distributions respectively,
where $f_a(x,Q_0^2)$ is the nuclear parton distribution which is taken from
Deep Inelastic Scattering experiments on nuclei at the relevant $Q_0^2$ and
$g(p_t)$ is parametrized by a Gaussian fit to proton--proton scattering data.
The spatial distribution of the partons is described by a convolution of a
Woods--Saxon distribution of nucleons in the nucleus and an exponential
distribution of individual partons within each nucleon. Details regarding both
initial conditions may be found in Ref.~\cite{GeiRep}.
Another model which takes as input the perturbative QCD cross sections is the
HIJING model~\cite{Wang,Gyul} which describes nuclear scattering in an eikonal 
formalism which convolves binary nucleon collisions. In both models, detailed
predictions have been made for various observables at RHIC--in particular, for
mini--jet production.

As suggested by the above, both models make assumptions which are not
necessarily motivated by perturbative QCD. To an extent, this is inevitable
because one is forced to model the soft physics. Where these approaches differ
significantly from the Weizs\"acker--Williams approach is in the factorization
assumption, namely, that partons from one nucleus resolve individual partons of
the other in each hard scattering. We have argued that the small $x$ partons
which dominate the physics of the central region instead have highly coherent
``wave-- like" interactions. This results in a vastly different space--time 
picture for the nuclear collision--at least for the very primordial stage of
the nuclear collision.

Naturally, the predictions of these models will differ significantly from the
Weizs\"acker--Williams model. For instance, because of the intrinsic $p_t\sim
\mu$ carried by the Weizs\"acker--Williams (or ``equivalent") gluons, gluon
production is enhanced by a factor $\alpha_s$ relative to the lowest order
$gg\rightarrow gg$ mini--jet process in a cascade. A simple explanation for
this enhancement is that because the valence quarks absorb the ``recoil", two
off--shell equivalent gluons can combine to produce an on--shell gluon. This
will impact significantly the many signatures to be studied at RHIC and LHC
such as jet production and dilepton and photon production. Further, the
intrinsic $p_t$ of the gluons ensures that intrinsic charm and strangeness
production is significantly larger in the Weizs\"acker--Williams
model~\cite{LV3}.

The non--perturbative models~\cite{String} primarily attempt to describe the
soft physics in ultrarelativistic nuclear collisions so it is not clear that
there is much overlap with the Weizs\"acker--Williams model. However, the
latter does provide some insight into one of these approaches, which we shall
dub the ``color capacitor" approach. Here it is assumed that the nuclei produce
a homogeneous chromo--electric field which produces particles
non--perturbatively by a mechanism analogous to the Schwinger mechanism for
strong electromagnetic fields. The evolution of these fields (including
back-reaction) is determined by a Boltzmann--like equation where the source
term now is given by the pair production rate~\cite{Eisenberg,Kluger}.

An important assumption in these color capacitor models is that of homogeneity
of the initial field configurations. However, the results discussed in the
previous section suggest that the Yang--Mills fields are highly non--linear and
inhomogeneous. The time scale $\tau>>1/\alpha_s\mu$ is the time scale for the
dissipation of the non--linearities in the fields. It would be interesting to
see how the solutions to the transport equations are modified  for initial
conditions given by the inhomogeneous Weizs\"acker--Williams field
configurations.

\section{Conclusions}
\vspace*{0.3cm}

We have described in this Comment a QCD based approach to describe the initial
conditions for ultrarelativistic nuclear collisions. The central region of
these collisions is dominated by ``wee" partons which carry only a small
fraction of the nuclear momentum. We have argued that for very large nuclei
these partons are only weakly coupled to each other. However, due to their
large density, many body effects are extremely important. The classical
behavior of these quanta (the QCD analogue of the Weizs\"acker--Williams
equivalent photons) can be described by an effective two dimensional field
theory. Quantum effects are treated by constructing the small fluctuations
propagator in the background field of these quanta.

An important objective of this approach is to understand if there is a
``Lipatov region" in nuclei where the parton densities grow rapidly and if the
shadowing of parton distributions in nuclei can be understood to result from
the precocious onset of parton screening. It is probable that deep inelastic
scattering experiments off large nuclei will be performed at HERA in the near
future~\cite{Strikman}. If so, one may expect unprecedentedly high parton
densities and interesting and perhaps unexpected phenomena in these
experiments.

These DIS experiments on nuclei at HERA would nicely complement the heavy ion
program at RHIC and especially LHC since they probe the same range of Bjorken
$x$. The results of these experiments would therefore place strong bounds on
mini--jet multiplicities and other signatures of nuclear collisions. Note that
these observables are extremely sensitive to the initial parton distributions
(for a discussion, see Ref.~\cite{Eskola}). However, to fully understand the
dynamics of nuclear collisions at central rapidities, we have to understand the
initial conditions ab initio--preferably in a QCD based approach like the one
discussed in this paper.

At the moment there are still many open questions which remain unresolved. An
empirical question is with regard to the applicability of weak coupling methods
to large nuclei. Is the bare parameter $\mu^2\sim A^{1/3}$ fm$^{-2}$ large
enough? One may argue on the basis of Renormalization Group arguments that this
parameter should effectively be larger and should grow with the increasing
parton density at small $x$. However, these arguments are not rigorous at this
stage.

A more serious problem is suggested by recent lattice simulations of the 2--D
effective field theory which show that the classical correlation functions
diverge quadratically with the lattice size $L$. Identifying why this
divergence occurs and how the background field may be modified accordingly
needs to be resolved satisfactorily. Finally, the problem of Lipatov
enhancement and saturation in nuclear parton distributions is not yet settled.

Despite the many technical problems that remain, there is much cause for
optimism since it appears now that the problem of initial conditions in
ultrarelativistic nuclear collisions can be treated systematically in a QCD
based approach. Because the various empirical signatures depend sensitively on
the initial conditions, one may hope to identify and interpret the elusive
quark gluon plasma in ultrarelativistic nuclear collisions at RHIC and LHC
early in the next millenium.

\section*{Acknowledgments}

Foremost, I would like to acknowledge my debt to my collaborators A. Ayala, R.
Gavai, J. Jalilian--Marian, A. Kovner, H. Weigert and especially L. McLerran. I
would also like to acknowledge useful discussions with I. Balitsky, B.
M\"uller, R. Narayanan, J. Qiu, T. Schaefer and M. Strikman. Finally, I would
like to thank J. Kapusta for his patience and good humor. 

Research supported by the U.S. Department of Energy under grants No. 
No. DOE Nuclear DE--FG06--90ER--40561.

\end{document}